\documentclass[twocolumn, deluxetables]{aastex631}
\usepackage{amsmath}
\usepackage{float}
\usepackage{multirow}
\usepackage{booktabs}
\usepackage{graphicx}
\usepackage{upgreek}

\graphicspath{{./}{figures/}}

\shorttitle{Swift Monitoring of NGC 1275}
\shortauthors{Ketchum et al.}

\begin{document}

\title{X-ray Flaring and Variability in NGC 1275, the Heart of the Perseus Cluster}

\author[0009-0001-4625-9240] {Sarah Ketchum}
\affiliation{Department of Astronomy, University of Michigan, Ann Arbor, MI 48109, USA}
\email{ketchum@umich.edu}

\author[0000-0003-2869-7682]{Jon M. Miller}
\affiliation{Department of Astronomy, University of Michigan, Ann Arbor, MI 48109, USA}
\email{jonmm@umich.edu}

\author[0000-0003-0667-5941]{Steven W. Allen}
\affiliation{Department of Physics, Stanford University, Stanford, CA 94305, USA}

\author[0000-0002-3687-6552]{Doyee Byun}
\affiliation{Department of Astronomy, University of Michigan, Ann Arbor, MI 48109, USA}

\author{Tianyin Hu}
\affiliation{Department of Astronomy, University of Michigan, Ann Arbor, MI 48109, USA}

\author[0000-0002-4992-4664]{Missagh Mehdipour}
\affiliation{Department of Astronomy, University of Michigan, Ann Arbor, MI 48109, USA}

\author{Veronica Tananko}
\affiliation{Department of Astronomy, University of Michigan, Ann Arbor, MI 48109, USA}

\author[0000-0002-7129-4654]{Xin Xiang}
\affiliation{Department of Astronomy, University of Michigan, Ann Arbor, MI 48109, USA}

\author[0000-0001-7630-8085]{Irina Zhuravleva}
\affiliation{Department of Astronomy \& Astrophysics, University of Chicago, Chicago, IL  60637, USA}




\begin{abstract}
NGC 1275 is the central galaxy in the Perseus Cluster. The active galactic nucleus (AGN) within NGC 1275 is notable for its strong and variable radio activity, tied to the production of radio jets that inflate large bubbles in the hot intracluster medium (ICM).  High spatial resolution X-ray imaging can separate the AGN from the bright ICM, but monitoring the mass accretion rate onto the black hole and establishing disk--jet connections in NGC~1275 requires a high cadence. 
Here, we report on X-ray monitoring of NGC~1275 using data taken over 20 years with the Neil Gehrels Swift Observatory.  Modeling the temporally constant ICM in each observation allows X-ray emission from accretion onto the black hole to be traced reliably, with typical flux errors of $\sim 3\%$.  X-ray flaring by a factor of $\sim2$ over mere days is detected starting on MJD 59956 (2023 Feb. 21).  The flares imply an emission region consistent with $r \leq 870~(10^{8}~M_{\odot}/M_{BH})~ GM/c^{2}$.   The profile of the flaring is inconsistent with simple predictions for tidal disruption events. A flare appears roughly 300 days later in radio monitoring data at 43~GHz.  Overall, our results indicate that coordinated, moderate-resolution X-ray imaging and radio monitoring could potentially trace disk--jet connections in the AGN that most vividly impact large-scale structure, and be extended to other sources that impact their hosts.
\end{abstract}


\section{Introduction}
The dominant galaxies at the centers of massive galaxy groups and clusters represent the pinnacles of galaxy formation. They are also among the clearest examples of how feedback processes from supermassive black holes limit galaxy growth. NGC 1275 ($z=0.01756$) is the dominant galaxy at the center of the Perseus Cluster, the X-ray brightest galaxy cluster in the sky \citep{Salome2011}. The galaxy itself has a very complex structure, with emission-line filaments that are likely caused by cool gas from the galaxy center following in the wake of buoyantly rising radio bubbles \citep{Fabian2008}. Additionally, there is a high-velocity component merging with the primary, low-velocity part of the galaxy \citep{Conseliceetal2001}. 

In radio bands, NGC~1275 is better known as 3C~84, a radio source with a flux density of several Jy at typical wavelengths (see, e.g., \citealt{lister2018}).  In concert with a combination of local gas and dust obscuration, NGC~1275 is often described as a Type-2 radio galaxy \citep{Paraschosetal2022}, or sometimes as a Type-2 Seyfert galaxy (e.g., \citealt{Liodakisetal2024}, \citealt{KhachikianWeedman1974}).  Other work has characterized NGC~1275 as a Seyfert-1.5 galaxy, potentially indicating that the central engine is not fully obscured \citep{Hoetal1997}.  

The approaching radio jet is viewed at an inclination of $\theta = 65^\circ \pm 16^\circ$ \citep{FujitaNagai2017}. However, the jet in 3C 84 is known to bend and change its launching direction in an irregular pattern \citep{2025A&A...696A..17F}. Nonetheless, according to measurements of this jet and the assumption that the disk is perpendicular to the jet, the disk inclination is estimated to be $45^\circ \pm 10^\circ$ \citep{Scharwachter2013}. This inclination is qualitatively consistent with a Seyfert-1.5 line of sight that may afford a view of the central engine.  Moreover, studies of this jet also imply that the AGN features an inhomogeneous and sub-parsec-scale accretion disk with a density of at least $10^{15}$ cm$^{-3}$ \citep{FujitaNagai2017}. 

Recent studies of radio, X-ray, and $\gamma$-ray emission in NGC~1275 reveal complex relationships between these bands (e.g., \citealt{fukazawa2018}, \citealt{dutson2014}).  Between 2013--2015, for instance, the X-ray and $\gamma$-ray emission were found to be correlated, potentially indicating that non-thermal radiation is produced downstream in the jet \citep{fukazawa2018}.  However, observations with XMM-Newton reveal a neutral Fe~K$_{\alpha}$ emission line in NGC~1275 \citep{churazov2003}; this line was later confirmed Suzaku and Hitomi (\citealt{yamazaki2013}, \citealt{hitomi2016}).  Chandra imaging confines cold Fe~K$_{\alpha}$ emission from the accretion flow to be less than 100~pc in extent \citep{miller2017}.  This suggests that the X-ray emission in NGC~1275 is not beamed, but is instead produced close to the black hole and seen by the accretion flow.  This may be supported by the moderate inclination of the system.  It is possible that typical X-ray emission or even X-ray flares without radio and $\gamma$-ray counterparts are produced close to the black hole, while X-ray emission associated with flares is produced downstream in the jet.  

The radio jet launched by the black hole in NGC~1275 appears to shape the hot X-ray gas within the core of the Perseus cluster, by inflating bubbles within the hot plasma (e.g., \citealt{fabian2003}).  NGC~1275 is just one of many AGN that shape their environment on a range of scales (for a review, see \citealt{fabian2012}), but the high X-ray surface brightness of the Perseus cluster reveals the interaction in unusual detail.  Efforts to understand disk--jet coupling have paired X-ray and radio observations across the black hole mass scale.  In the stellar-mass black hole GRS~1915$+$105 and quasar 3C~120, for instance, radio flares are sometimes preceded by a dip in X-ray flux, suggesting the ejection of the innermost accretion flow (see, e.g., \citep{chat2009}).  Although Chandra can separate the AGN from the diffuse ICM emission, Chandra cannot monitor NGC~1275 at a high cadence, inhibiting efforts to study disk--jet coupling in what is arguably the most compelling setting.

The stellar mass of NGC~1275 has been measured as $M_{\star} = 2.43 \times 10^{11} M_{\odot}$\citep{Mathewsetal2006}, but the mass of the black hole remains a matter of debate.  \citet{Scharwachter2013} used data simulated by Gemini Near-Infrared Integral Field Spectrograph (NIFS) cubes in order to estimate the mass enclosed within the H2 disk of the galaxy. They found the black hole mass to be $8^{+7}_{-2} \times10^{8}~M_{\odot}$, with the molecular gas of the central region of the galaxy (R $<$ 50 pc) estimated to have a mass of $4 \times 10^{8}~M_{\odot}$. This relation matches the M- $\sigma$ relationship within $rms$ scatter, but due to the high uncertainty created by mass of the molecular gas within the center of the galaxy, this number might be better interpreted as an upper limit on the black hole mass \citep{Scharwachter2013}.  \citet{Riffeletal2020} used NIFS to analyze CO bandheads in order to find a stellar dispersion velocity of $\sigma = 265 \pm 26$ km/s. The stellar dispersion velocity was used to find a supermassive black hole mass estimate of $1.1^{+0.9}_{-0.5} \times 10^{9}~M_{\odot}$. This agrees with dynamical measurements.   In contrast, \citet{Onorietal2017} arrived at a black hole mass of $M = 2.9 \times 10^{7}~M_{\odot}$ based on the Pa$\beta$ line width and scaling relation.  It should be noted that this low mass falls far below the $M-\sigma$ relationship.

In this paper, we present a study of nearly 20~years of Swift/XRT monitoring of NGC~1275.  Section 2 describes how the data were reduced.  Our analysis and results are described in Section 3.  Finally, Section 4 discusses the implications of our findings, and the need for coordinated, high-cadence, multi-wavelength monitoring of NGC~1275 and similar AGN. 

\section{Observations and Data Reduction}
We utilize every Swift/XRT observation of NGC~1275 obtained in ``photon counting'' mode before March 2026.  We omitted data obtained in ``windowed timing'' mode because the loss of imaging prevents effective characterization and subtraction of the ICM emission.  In total, then, we consider 89 observations. 
All of the data were processed using HEASOFT version 6.34, and the corresponding calibration files for this release.

Spectral files were created by extracting events using a circular region with a radius of 18 arc seconds, centered on the position of the X-ray nucleus of NGC~1275.  Background spectra were not extracted as the diffuse emission in the Perseus Cluster varies continually with radius, and our goal is to characterize relative flux levels in NGC~1275.  The \texttt{xrtmkarf} tool was then used to create an exposure-corrected ancillary response file (arf) for each observation.  In all cases, the most recent standard ``photon counting'' redistribution matrix file (rmf) file was used in spectral fitting.  

In order to characterize diffuse cluster emission within our extraction region, we summed the first 40 spectra and responses using the tools \texttt{addascaspec} and \texttt{ftaddrmf}.  This selection avoids previously reported flares and those reported in this work.  The summed spectrum was fit first, and the diffuse emission parameters were then held fixed in subsequent fits to individual observations, since the cluster emission is not expected to vary.

To maximize the sensitivity of each individual spectrum, the data were ``optimally binned'' using the tool \texttt{ftgrouppha} (see \citealt{kaastra2016}).  Spectral fitting was restricted to the 0.3--10.0~keV band over which the XRT calibration holds in ``photon counting'' mode.  All spectral fits were made in XSPEC version 12.14.1 \citep{arnaud1996}, minimizing a Cash statistic \citep{cash1979}.  All of the uncertainties reported in this work are $1\sigma$ errors.

\section{Analysis and Results} 
In order to characterize the ICM within the Perseus cluster, we initially made fits to the summed Swift/XRT spectrum.   We adopted a simple model consisting of diffuse plasma emission from the ICM, and power-law emission from accretion on to the massive black hole.  Within XSPEC, the model can be written as \texttt{tbabs*zmshift*(apec+pow)}.   The \texttt{tbabs} component accounted for neutral foreground absorption along the line of sight through a variable column density, ${\rm N}_{\rm H}$.  The \texttt{zmshift} component was used to shift the total emission model to the frame of NGC~1275 ($z=0.01756$) and included no free parameters. The \texttt{apec} component modeled diffuse ICM emission \citep{foster2012}.  We measured the plasma temperature and flux normalization, but kept other parameters (e.g., the metallicity) frozen at the default solar values as defined in \citet{2000ApJ...542..914W}.  In preliminary fits to the summed spectrum, the power-law index is measured to be $\Gamma = 1.68^{+0.06}_{-0.04}$.  This is fully consistent with the $\Gamma = 1.7$ index typical of bright Seyfert AGN \citep{nandra2007}, so we fixed a value of $\Gamma = 1.7$ while allowing the power-law normalization to vary freely.

This model did not result in a statistically acceptable fit (Cash$=79$ for $\nu = 61$ degrees of freedom). The strongest deviations are $\leq$10\% residuals in the Si~K band (roughly 1.7--2.1~keV).  We measure a column density of ${\rm N}_{\rm H} = 1.99\pm 0.04\times 10^{21}~{\rm cm}^{-2}$, a plasma temperature of $kT = 5.2\pm 0.1$~keV, and a flux normalization of $K = 8.2\pm 1.6\times 10^{-3}$ (a normalized emission measure with physical units of ${\rm cm}^{-5}$).  

In the spectra obtained with from Hitomi and XRISM (for instance), the ICM within Perseus requires temperature gradients, velocity gradients, abundance variations, and other complexities (\citealt{hitomi2016}, \citealt{xrism2026}).   However, our simple model provides a basic characterization of the strongest emission lines and continuum.  In particular, it fits the 6.7~keV line from Fe~XXV that might otherwise bias determinations of the power-law flux from NGC~1275.

In subsequent spectral fits to individual observations, we froze the best-fit plasma temperature and flux normalization values measured in the summed spectrum (as well as the best-fit neutral column density along the line of sight), and measured variations in the power-law flux normalization.  These simple fits yielded acceptable fits.  The total unabsorbed power-law flux and errors in the 0.3--10.0~keV band were measured by removing the line of sight absorber.

Table 1 lists the total unabsorbed flux in the 0.3-10.0~keV band for each Swift/XRT observation of NGC 1275. We infer a mean unabsorbed flux of $F = 5.40 \pm 0.3 \times 10^{-11}~{\rm erg}~{\rm cm}^{-2}~{\rm s}^{-1}$. The median unabsorbed flux is $F = 4.94 \pm 0.2 \times 10^{-11}~{\rm erg}~{\rm cm}^{-2}~{\rm s}^{-1}$. NGC~1275 has a redshift of z=0.01756, corresponding to a luminosity distance of approximately 76.3 Mpc, assuming a standard flat $\Lambda$CDM cosmology with the default parameter values adopted in XSPEC \citep{arnaud1996}: $\Omega_{\rm M}=0.27$, $\Omega_{\Lambda}=0.73$, and $H_0=70~{\rm km~s^{-1}~Mpc^{-1}}$. Using $L=4\pi D_L^2F$, these fluxes correspond to a mean X-ray luminosity of $L_{\rm mean}=3.76\pm0.21\times10^{43}~{\rm erg~s^{-1}}$ and a median X-ray luminosity of $L_{\rm median}=3.44\pm0.14\times10^{43}~{\rm erg~s^{-1}}$.

Figures 1 shows the Swift/XRT light curve of NGC~1275, over the full monitoring period.  There are large time gaps between some observations, particularly between MJD 58455 and 59931 (nearly 1500 days).  The mean and median fluxes are plotted so that individual points may be compared to these key values.  Despite the modest sensitivity of individual spectra, many individual points differ significantly from the long-term mean and median values.  The most striking feature of the light curve is the strong flaring activity just prior to MJD~60000 (highlighted in pink in Figure 1).  In rough terms, the flux of NGC~1275 jumped by a factor of $\sim2.0$.  We note that \cite{fukazawa2018} studied X-ray and gamma-ray variations in NGC~1275, but only prior to MJD 57100.

In order to verify that the flare and our flux determinations are robust, we explicitly compared the observation with the highest total flux (ObsID 34765023, obtained on MJD 59956.1), to the summed spectrum.  The two spectra are shown in Figure 2.  Visually, it is apparent that the two spectra only differ by an absorbed power-law component, attributable to accretion onto the massive black hole. 

Figure 2 also shows a fit to the difference spectrum (high flux observation minus summed spectrum).  In order to ensure proper subtraction, the difference spectrum was fit using SPEX version 3.08.01 \citep{kaastra1996}, by reading the high flux spectrum in as data and the summed spectrum as background (avoiding potential numerical artifacts when using the HEASARC suite).  First binning using the ``optimal'' algorithm and secondarily by a factor of eight to deliver a number of bins similar to individual observations, an absorbed $\Gamma = 1.7$ power-law gives a Cash statistic of $C = 97.3$ for $\nu = 94$ degrees of freedom, signaling an acceptable fit.

Figure 3 shows the Swift/XRT light curve during the flaring period.  In total, the newly detected activity appears to last less than 60 days, between MJD 59930 and 55990 (although additional flaring may have been missed by the sporadic monitoring; we comment on this in the next section). The flaring activity is likely composed of at least two strong X-ray peaks. 

Both peaks have characteristic durations of approximately $\sim5$~days, giving an upper limit on the emission radius of $r \leq c\Delta t \leq 7.8\times 10^{15}~{\rm cm}$, or $r \leq 870~ (10^{8}~M_{\odot}/M_{BH})~ GM/c^{2}$.  The size scale implied by the observed flaring activity is easily consistent with a compact coronal region wherein magnetic flux may help to launch a jet, as per magnetically arrested disk modes (e.g., \citealt{sasha2011}).  This limit and the fact that the variable emission is consistent with a Seyfert-like power-law index of $\Gamma = 1.7$ suggest that the newly detected flares may be related to accretion activity (X-ray spectra of blazars measure $\Gamma = 2.0-2.4$; see, e.g., \citealt{saad2021}).  However, the data do not definitively rule out an origin in downstream processes within the jet.

We therefore examined whether the flares could be consistent with a stellar tidal disruption event (TDE).  Black holes with masses exceeding $M \geq 10^{8}~M_{\odot}$ are unable to disrupt typical main sequence stars, because their gravitation changes too slowly across the stellar diameter.  Some estimates of the mass of the black hole in NGC~1257 place it below this limit, so it is worth examining if the flares are consistent with the canonical $F \propto t^{-5/3}$ flux trend expected for tidal disruption events (\citealt{rees1988}; see \citealt{miller2015} for an example).  For simplicity, we restricted our examination to the larger peak (the pink band in Figure 3).  

Figure 4 shows the stronger peak within the flare, the best-fit exponential flux decay curve, and the anticipated $F \propto t^{-5/3}$ decay.  The best-fit decay model is $F\propto t^{-0.27}$, far slower than $F\propto t^{-5/3}$.  Figure 5 shows the later, slightly weaker peak within the flare, which had a best-fit decay model of $F\propto t^{-0.11}$. Even though the flares differ markedly from other variability over nearly 20 years of X-ray monitoring, it is unlikely that they are related to a TDE.  It is more likely that these flares result from variations in the mass accretion rate onto the black hole or downstream shocks or other processes.

Past radio monitoring of NGC~1275 enables a basic comparison of the X-ray and radio flux trends over nearly 20 years.  Figure 6 shows the Swift/XRT light curve compared to 15~GHz and 43 GHz obtained through the MOJAVE and VLBA-BU-BLAZAR programs, respectively \citep{lister2018, Weaver2022}. The fractional variability for both the X-ray and radio data was calculated using the method laid out by \citet{Vaughanetal2003}. During the course of the study, the fractional variability was found to be 0.2857 (28.57\%) for the X-ray observations, 0.2809 (28.09\%) for the 15 GHz radio, and 0.2883 (28.83\%) for the 43 GHz radio. Additionally, the fractional variability was calculated for a subset of the data that excluded the X-ray flare (55000-59000 MJD). For this subset, the fractional variability was 0.1817 (18.17\%) for the X-ray data, 0.1940 (19.4\%) for the 15 GHz radio data, and 0.2760 (27.6\%) for the 43 GHz radio data.  

The 15~GHz radio flux density from NGC~1275 is sparsely sampled after MJD 60000, making it difficult to associate any variations with the prior X-ray flare.  The 15~GHz generally trends upward for 1000 days following the X-ray flare, relative to the 2000 days before the flare.  The higher cadence of the 43~GHz VLBA flux density may offer more insights.  Flaring above 4~Jy starts on MJD 60230, or 296~days after the start of the newly detected X-ray flaring.  This could represent a propagation time from the X-ray corona surrounding the black hole to a radio-emitting region within the jet, or a propagation time between two points downstream in the jet.  With such coarse sampling in all bands, however, caution is warranted.

\section{Discussion}
We have analyzed Swift/XRT monitoring observations of NGC 1275, from 2007 to 2026.  The flux of the AGN was inferred found by fitting a composite spectral model to 89 observations, including a fixed plasma emission component from the diffuse ICM and a variable power-law component from the AGN.  We find that NGC 1275 is significantly variable, and we report the discovery of strongest X-ray flaring yet observed starting in February 2023.  The flares are not consistent with simple predictions for TDE decay curves, and more likely represent dissipation related to unusual fluctuations in the mass accretion rate onto the massive black hole or downstream shocks in the jet.  In this section, we briefly discuss the consequences of our findings for our understanding of NGC~1275, and future studies of both this AGN and the Perseus cluster.

The Eddington limit for a black hole with a mass of $M = 10^{7}~M_{\odot}$ is $L = 1.3\times 10^{45}~{\rm erg}~{\rm s}^{-1}$.  This is about 40 times larger than the mean X-ray luminosity that we infer from 89 monitoring observations, $L = 3.5\times 10^{43}~{\rm erg}~{\rm s}^{-1}$. However, massive black holes release most of their energy in UV light. Extrapolating from known bolometric corrections for AGN of this size, we can assume that the true bolometric luminosity of the AGN in NGC~1275 is at least an order of magnitude higher \citep{2007MNRAS.381.1235V}. Therefore, $(L_{\rm bol}/L_{\rm Edd})$ is roughly 0.3, if the black hole mass is $M = 10^{7}~M_{\odot}$.  Requiring a lower Eddington fraction -- as is often associated with jet-producing systems -- would favor a larger black-hole mass, potentially $(M\gtrsim10^{8-9}~M_\odot)$. 

Prior work found that NGC~1275 is variable in X-rays \citep{fukazawa2018}, but the strong X-ray flaring that we detected just prior to MJD 60000 is new and of special interest.  Compared to typical flux variability of roughly 10\%, the flares represent nearly 100\% variations on time scales shorter than \textbf{5} days (each).  The flares are unlikely to be powered by stellar tidal disruption events, based on their markedly slower decay trends.

Three lines of circumstantial evidence may favor an accretion origin for the observed X-ray flares.  First, the variable emission in NGC~1275 is consistent with a $\Gamma = 1.7$ power-law, similar to Seyfert AGN wherein the central engine is visible.  Second, isotropic X-ray emission is needed to generate the narrow Fe~K$\upalpha$ emission line observed in NGC~1275 (e.g., \citealt{hitomi2016}, \cite{miller2017}).  Third, the characteristic flaring timescale of 5 days corresponds to an emission region of $r \leq 870~ (10^{8}~M_{\odot}/M_{BH})~ GM/c^{2}$, consistent with a compact X-ray corona that could illuminate the accretion flow and stimulate the Fe~K$\alpha$ line.  However, this does not mean that 100\% of the X-ray emission observed from NGC~1275 originates in a compact corona close to the black hole.  Some flares -- particularly those with contemporaneous increases in radio and $\gamma$-ray emission -- could arise downstream in the relativistic jet.   We cannot exclude that the newly detected flares also originate downstream in the jet.

Despite the prior lack of dense, simultaneous X-ray and radio observations of NGC~1275, prior flux trends offer some tantalizing hints of similarities to other AGN.  The radio points that most closely precede the strong X-ray flares lie below the mean and median values.  In a diverse set of black holes, including the AGN NGC~4051 and 3C~120, and the stellar-mass black holes GRS~1915$+$105 and Cygnus X-1, {\em inverse} correlations are found between X-ray and radio flux trends (\citealt{king2011}; also see \citep{chat2009}).   There is tentative evidence that inverse coupling may take hold above an Eddington fraction of 0.1; this would imply that the mass of the black hole in NGC~1275 may not greatly exceed $M \sim 10^{9}~M_{\odot}$.  The propagation time from the base of a jet to the radio-emitting regions is largely unknown and may depend on many factors; it is possible that the points above 4~Jy in the 43~GHz VLBA light curve signal a $\sim300$ day propagation time between a compact X-ray corona and radio-emitting region, or between two regions downstream in the jet.

The long-term X-ray light curve shown in Figure 1 and Figure 6 is sparsely and erratically sampled, with only one observation in the MJD 58000-59900 window (for instance).  The sampling across the newly discovered flare intervals is also sparse and irregular.  Coordinated, higher-cadence monitoring of NGC~1275 in X-ray and radio bands -- particularly if it affords a view of the central engine -- could reveal a great deal more about this source and similar black holes.  A systematic program of daily observations over a period of years could reveal the true variability fraction in X-rays, the duty cycle of flaring, structure within flares, and the characteristics of flare decays.  It will then be possible to determine the number of X-ray and radio-flares that lack an immediate counterpart, and to search for lags that could confirm or reject the $\sim300$~day lag that may be indicated in current data.

Our results may also illustrate a direction for optimizing future studies of feedback within clusters.  The half-power diameter of XRISM mirror systems is approximately 1.7 arc minutes \citep{tashiro2025}, and this is sampled by pixels that are 30 arc seconds on each side.  During periods when the AGN is bright, the central pixels contain relatively more AGN flux.  This affects spatial analysis of the cluster emission, and diminishes the power of line diagnostics of the cluster core gas by adding a heightened continuum.  In Perseus, for instance, the ratio of turbulent to thermal pressure in the innermost 20~kpc is just 4--6\% (\citealt{hitomi2016}, \citealt{xrism2026});  although line velocity widths are minimally impacted by AGN activity, abundance measurements are impacted.  Coordinated, high-cadence monitoring of cluster AGN with Swift (for instance) could help to identify periods of low AGN activity that enable improved plasma diagnostics in cluster cores and better AGN feedback constraints.

We thank the anonymous referee for comments that improved this manuscript.  SK acknowledges support from University of Michigan MRADS (Michigan Research and Discovery Scholars). JMM acknowledges support from XRISM through NASA/GSFC. This study makes use of VLBA data from the VLBA-BU Blazar Monitoring Program (BEAM-ME and VLBA-BU-BLAZAR; http://www.bu.edu/blazars/BEAM-ME.html), funded by NASA through the Fermi Guest Investigator Program. The VLBA is an instrument of the National Radio Astronomy Observatory. The National Radio Astronomy Observatory is a facility of the National Science Foundation operated by Associated Universities, Inc. We acknowledge the use of public data from the Swift data archive.

\bibliography{main}{}

\begin{thebibliography}{}
\expandafter\ifx\csname natexlab\endcsname\relax\def\natexlab#1{#1}\fi
\providecommand{\url}[1]{\href{#1}{#1}}
\providecommand{\dodoi}[1]{doi:~\href{http://doi.org/#1}{\nolinkurl{#1}}}
\providecommand{\doeprint}[1]{\href{http://ascl.net/#1}{\nolinkurl{http://ascl.net/#1}}}
\providecommand{\doarXiv}[1]{\href{https://arxiv.org/abs/#1}{\nolinkurl{https://arxiv.org/abs/#1}}}

\bibitem[{{Arnaud}(1996)}]{arnaud1996}
{Arnaud}, K.~A. 1996, in Astronomical Society of the Pacific Conference Series,
  Vol. 101, Astronomical Data Analysis Software and Systems V, ed. G.~H.
  {Jacoby} \& J.~{Barnes}, 17

\bibitem[{{Cash}(1979)}]{cash1979}
{Cash}, W. 1979, \apj, 228, 939, \dodoi{10.1086/156922}

\bibitem[{{Chatterjee} {et~al.}(2009){Chatterjee}, {Marscher}, {Jorstad},
  {Olmstead}, {McHardy}, {Aller}, {Aller}, {L{\"a}hteenm{\"a}ki}, {Tornikoski},
  {Hovatta}, {Marshall}, {Miller}, {Ryle}, {Chicka}, {Benker}, {Bottorff},
  {Brokofsky}, {Campbell}, {Chonis}, {Gaskell}, {Gaynullina}, {Grankin},
  {Hedrick}, {Ibrahimov}, {Klimek}, {Kruse}, {Masatoshi}, {Miller}, {Pan},
  {Petersen}, {Peterson}, {Shen}, {Strel'nikov}, {Tao}, {Watkins}, \&
  {Wheeler}}]{chat2009}
{Chatterjee}, R., {Marscher}, A.~P., {Jorstad}, S.~G., {et~al.} 2009, \apj,
  704, 1689, \dodoi{10.1088/0004-637X/704/2/1689}

\bibitem[{{Churazov} {et~al.}(2003){Churazov}, {Forman}, {Jones}, \&
  {B{\"o}hringer}}]{churazov2003}
{Churazov}, E., {Forman}, W., {Jones}, C., \& {B{\"o}hringer}, H. 2003, \apj,
  590, 225, \dodoi{10.1086/374923}

\bibitem[{{Conselice} {et~al.}(2001){Conselice}, {Gallagher}, \&
  {Wyse}}]{Conseliceetal2001}
{Conselice}, C.~J., {Gallagher}, John~S., I., \& {Wyse}, R. F.~G. 2001, The
  Astronomical Journal, 122, 2281, \dodoi{10.1086/323534}

\bibitem[{{Dutson} {et~al.}(2014){Dutson}, {Edge}, {Hinton}, {Hogan},
  {Gurwell}, \& {Alston}}]{dutson2014}
{Dutson}, K.~L., {Edge}, A.~C., {Hinton}, J.~A., {et~al.} 2014, \mnras, 442,
  2048, \dodoi{10.1093/mnras/stu975}

\bibitem[{{Fabian}(2012)}]{fabian2012}
{Fabian}, A.~C. 2012, \araa, 50, 455,
  \dodoi{10.1146/annurev-astro-081811-125521}

\bibitem[{{Fabian} {et~al.}(2008){Fabian}, {Johnstone}, {Sanders}, {Conselice},
  {Crawford}, {Gallagher}, \& {Zweibel}}]{Fabian2008}
{Fabian}, A.~C., {Johnstone}, R.~M., {Sanders}, J.~S., {et~al.} 2008, Nature,
  454, 968, \dodoi{10.1038/nature07169}

\bibitem[{{Fabian} {et~al.}(2003){Fabian}, {Sanders}, {Allen}, {Crawford},
  {Iwasawa}, {Johnstone}, {Schmidt}, \& {Taylor}}]{fabian2003}
{Fabian}, A.~C., {Sanders}, J.~S., {Allen}, S.~W., {et~al.} 2003, \mnras, 344,
  L43, \dodoi{10.1046/j.1365-8711.2003.06902.x}

\bibitem[{{Foschi} {et~al.}(2025){Foschi}, {G{\'o}mez}, {Fuentes}, {Cho},
  {Marscher}, \& {Jorstad}}]{2025A&A...696A..17F}
{Foschi}, M., {G{\'o}mez}, J.~L., {Fuentes}, A., {et~al.} 2025, \aap, 696, A17,
  \dodoi{10.1051/0004-6361/202453406}

\bibitem[{{Foster} {et~al.}(2012){Foster}, {Ji}, {Smith}, \&
  {Brickhouse}}]{foster2012}
{Foster}, A.~R., {Ji}, L., {Smith}, R.~K., \& {Brickhouse}, N.~S. 2012, \apj,
  756, 128, \dodoi{10.1088/0004-637X/756/2/128}

\bibitem[{{Fujita} \& {Nagai}(2017)}]{FujitaNagai2017}
{Fujita}, Y., \& {Nagai}, H. 2017, Monthly Notices of the Royal Astronomical
  Society, 465, L94, \dodoi{10.1093/mnrasl/slw221}

\bibitem[{{Fukazawa} {et~al.}(2018){Fukazawa}, {Shiki}, {Tanaka}, {Itoh},
  {Takahashi}, {Imazato}, {D'Ammando}, {Ojha}, \& {Nagai}}]{fukazawa2018}
{Fukazawa}, Y., {Shiki}, K., {Tanaka}, Y., {et~al.} 2018, \apj, 855, 93,
  \dodoi{10.3847/1538-4357/aaabc0}

\bibitem[{{Hitomi Collaboration} {et~al.}(2016){Hitomi Collaboration},
  {Aharonian}, {Akamatsu}, {Akimoto}, {Allen}, {Anabuki}, {Angelini}, {Arnaud},
  {Audard}, {Awaki}, {Axelsson}, {Bamba}, {Bautz}, {Blandford}, {Brenneman},
  {Brown}, {Bulbul}, {Cackett}, {Chernyakova}, {Chiao}, {Coppi}, {Costantini},
  {de Plaa}, {den Herder}, {Done}, {Dotani}, {Ebisawa}, {Eckart}, {Enoto},
  {Ezoe}, {Fabian}, {Ferrigno}, {Foster}, {Fujimoto}, {Fukazawa}, {Furuzawa},
  {Galeazzi}, {Gallo}, {Gandhi}, {Giustini}, {Goldwurm}, {Gu}, {Guainazzi},
  {Haba}, {Hagino}, {Hamaguchi}, {Harrus}, {Hatsukade}, {Hayashi}, {Hayashi},
  {Hayashida}, {Hiraga}, {Hornschemeier}, {Hoshino}, {Hughes}, {Iizuka},
  {Inoue}, {Inoue}, {Ishibashi}, {Ishida}, {Ishikawa}, {Ishisaki}, {Itoh},
  {Iyomoto}, {Kaastra}, {Kallman}, {Kamae}, {Kara}, {Kataoka}, {Katsuda},
  {Katsuta}, {Kawaharada}, {Kawai}, {Kelley}, {Khangulyan}, {Kilbourne},
  {King}, {Kitaguchi}, {Kitamoto}, {Kitayama}, {Kohmura}, {Kokubun}, {Koyama},
  {Koyama}, {Kretschmar}, {Krimm}, {Kubota}, {Kunieda}, {Laurent}, {Lebrun},
  {Lee}, {Leutenegger}, {Limousin}, {Loewenstein}, {Long}, {Lumb}, {Madejski},
  {Maeda}, {Maier}, {Makishima}, {Markevitch}, {Matsumoto}, {Matsushita},
  {McCammon}, {McNamara}, {Mehdipour}, {Miller}, {Miller}, {Mineshige},
  {Mitsuda}, {Mitsuishi}, {Miyazawa}, {Mizuno}, {Mori}, {Mori}, {Moseley},
  {Mukai}, {Murakami}, {Murakami}, {Mushotzky}, {Nagino}, {Nakagawa},
  {Nakajima}, {Nakamori}, {Nakano}, {Nakashima}, {Nakazawa}, {Nobukawa},
  {Noda}, {Nomachi}, {O'Dell}, {Odaka}, {Ohashi}, {Ohno}, {Okajima}, {Ota},
  {Ozaki}, {Paerels}, {Paltani}, {Parmar}, {Petre}, {Pinto}, {Pohl}, {Porter},
  {Pottschmidt}, {Ramsey}, {Reynolds}, {Russell}, {Safi-Harb}, {Saito},
  {Sakai}, {Sameshima}, {Sato}, {Sato}, {Sato}, {Sawada}, {Schartel},
  {Serlemitsos}, {Seta}, {Shidatsu}, {Simionescu}, {Smith}, {Soong}, {Stawarz},
  {Sugawara}, {Sugita}, {Szymkowiak}, {Tajima}, {Takahashi}, {Takahashi},
  {Takeda}, {Takei}, {Tamagawa}, {Tamura}, {Tamura}, {Tanaka}, {Tanaka},
  {Tanaka}, {Tashiro}, {Tawara}, {Terada}, {Terashima}, {Tombesi}, {Tomida},
  {Tsuboi}, {Tsujimoto}, {Tsunemi}, {Tsuru}, {Uchida}, {Uchiyama}, {Uchiyama},
  {Ueda}, {Ueda}, {Ueno}, {Uno}, {Urry}, {Ursino}, {de Vries}, {Watanabe}, \&
  {Werner}}]{hitomi2016}
{Hitomi Collaboration}, {Aharonian}, F., {Akamatsu}, H., {et~al.} 2016, \nat,
  535, 117, \dodoi{10.1038/nature18627}

\bibitem[{{Ho} {et~al.}(1997){Ho}, {Filippenko}, \& {Sargent}}]{Hoetal1997}
{Ho}, L.~C., {Filippenko}, A.~V., \& {Sargent}, W. L.~W. 1997, The
  Astrophysical Journal Supplement Series, 112, 315, \dodoi{10.1086/313041}

\bibitem[{{Kaastra} \& {Bleeker}(2016)}]{kaastra2016}
{Kaastra}, J.~S., \& {Bleeker}, J.~A.~M. 2016, \aap, 587, A151,
  \dodoi{10.1051/0004-6361/201527395}

\bibitem[{{Kaastra} {et~al.}(1996){Kaastra}, {Mewe}, \&
  {Nieuwenhuijzen}}]{kaastra1996}
{Kaastra}, J.~S., {Mewe}, R., \& {Nieuwenhuijzen}, H. 1996, in UV and X-ray
  Spectroscopy of Astrophysical and Laboratory Plasmas, ed. K.~{Yamashita} \&
  T.~{Watanabe}, 411--414

\bibitem[{{Khachikian} \& {Weedman}(1974)}]{KhachikianWeedman1974}
{Khachikian}, E.~Y., \& {Weedman}, D.~W. 1974, The Astrophysical Journal, 192,
  581, \dodoi{10.1086/153093}

\bibitem[{{King} {et~al.}(2011){King}, {Miller}, {Cackett}, {Fabian},
  {Markoff}, {Nowak}, {Rupen}, {G{\"u}ltekin}, \& {Reynolds}}]{king2011}
{King}, A.~L., {Miller}, J.~M., {Cackett}, E.~M., {et~al.} 2011, \apj, 729, 19,
  \dodoi{10.1088/0004-637X/729/1/19}

\bibitem[{{Liodakis} {et~al.}(2024){Liodakis}, {Chakraborty}, {Marin},
  {Ehlert}, {Barnouin}, {Kouch}, \& {others}}]{Liodakisetal2024}
{Liodakis}, I., {Chakraborty}, S., {Marin}, F., {et~al.} 2024, The
  Astrophysical Journal, 970, 41, \dodoi{10.3847/1538-4357/ad4d9a}

\bibitem[{{Lister} {et~al.}(2018){Lister}, {Aller}, {Aller}, {Hodge}, {Homan},
  {Kovalev}, {Pushkarev}, \& {Savolainen}}]{lister2018}
{Lister}, M.~L., {Aller}, M.~F., {Aller}, H.~D., {et~al.} 2018, \apjs, 234, 12,
  \dodoi{10.3847/1538-4365/aa9c44}

\bibitem[{{Mathews} {et~al.}(2006){Mathews}, {Faltenbacher}, \&
  {Brighenti}}]{Mathewsetal2006}
{Mathews}, W.~G., {Faltenbacher}, A., \& {Brighenti}, F. 2006, The
  Astrophysical Journal, 638, 659, \dodoi{10.1086/499119}

\bibitem[{{Miller} {et~al.}(2017){Miller}, {Bautz}, \& {McNamara}}]{miller2017}
{Miller}, J.~M., {Bautz}, M.~W., \& {McNamara}, B.~R. 2017, \apjl, 850, L3,
  \dodoi{10.3847/2041-8213/aa9566}

\bibitem[{{Miller} {et~al.}(2015){Miller}, {Kaastra}, {Miller}, {Reynolds},
  {Brown}, {Cenko}, {Drake}, {Gezari}, {Guillochon}, {Gultekin}, {Irwin},
  {Levan}, {Maitra}, {Maksym}, {Mushotzky}, {O'Brien}, {Paerels}, {de Plaa},
  {Ramirez-Ruiz}, {Strohmayer}, \& {Tanvir}}]{miller2015}
{Miller}, J.~M., {Kaastra}, J.~S., {Miller}, M.~C., {et~al.} 2015, \nat, 526,
  542, \dodoi{10.1038/nature15708}

\bibitem[{{Nandra} {et~al.}(2007){Nandra}, {O'Neill}, {George}, \&
  {Reeves}}]{nandra2007}
{Nandra}, K., {O'Neill}, P.~M., {George}, I.~M., \& {Reeves}, J.~N. 2007,
  \mnras, 382, 194, \dodoi{10.1111/j.1365-2966.2007.12331.x}

\bibitem[{{Onori} {et~al.}(2017){Onori}, {Ricci}, {La Franca}, {Bianchi},
  {Bongiorno}, {Brusa}, {Marconi}, {Onado}, \& {Sani}}]{Onorietal2017}
{Onori}, F., {Ricci}, F., {La Franca}, F., {et~al.} 2017, Monthly Notices of
  the Royal Astronomical Society: Letters, 468, L97,
  \dodoi{10.1093/mnrasl/slx032}

\bibitem[{{Paraschos} {et~al.}(2022){Paraschos}, {Krichbaum}, {Kim}, {Hodgson},
  {Oh}, {Ros}, {Zensus}, {Marscher}, {Jorstad}, {Gurwell},
  {L{\"a}hteenm{\"a}ki}, {Tornikoski}, {Kiehlmann}, \&
  {Readhead}}]{Paraschosetal2022}
{Paraschos}, G.~F., {Krichbaum}, T.~P., {Kim}, J.-Y., {et~al.} 2022, Astronomy
  \& Astrophysics, 665, A1, \dodoi{10.1051/0004-6361/202243343}

\bibitem[{{Rees}(1988)}]{rees1988}
{Rees}, M.~J. 1988, \nat, 333, 523, \dodoi{10.1038/333523a0}

\bibitem[{{Riffel} {et~al.}(2020){Riffel}, {Storchi-Bergmann}, {Zakamska}, \&
  {Riffel}}]{Riffeletal2020}
{Riffel}, R.~A., {Storchi-Bergmann}, T., {Zakamska}, N.~L., \& {Riffel}, R.
  2020, Monthly Notices of the Royal Astronomical Society, 496, 4857,
  \dodoi{10.1093/mnras/staa1922}

\bibitem[{{Saad} {et~al.}(2021){Saad}, {Nasser}, {Abdelbar}, \&
  {Beheary}}]{saad2021}
{Saad}, A.~A., {Nasser}, A.~M., {Abdelbar}, A.~M., \& {Beheary}, M.~M. 2021,
  \rmxaa, 57, 133, \dodoi{10.22201/ia.01851101p.2021.57.01.09}

\bibitem[{{Salom{\'e}} {et~al.}(2011){Salom{\'e}}, {Combes}, {Revaz}, {Downes},
  {Edge}, \& {Fabian}}]{Salome2011}
{Salom{\'e}}, P., {Combes}, F., {Revaz}, Y., {et~al.} 2011, \aap, 531, A85,
  \dodoi{10.1051/0004-6361/200811333}

\bibitem[{{Scharw{\"a}chter} {et~al.}(2013){Scharw{\"a}chter}, {McGregor},
  {Dopita}, \& {Beck}}]{Scharwachter2013}
{Scharw{\"a}chter}, J., {McGregor}, P.~J., {Dopita}, M.~A., \& {Beck}, T.~L.
  2013, Mon. Not. R. Astron. Soc., 429, 2315, \dodoi{10.1093/mnras/sts502}

\bibitem[{{Tashiro} {et~al.}(2025){Tashiro}, {Kelley}, {Watanabe}, {Maejima},
  {Reichenthal}, {Toda}, {Hartz}, {Santovincenzo}, {Matsushita}, {Yamaguchi},
  {Petre}, {Williams}, {Guainazzi}, {Costantini}, {Takei}, {Ishisaki},
  {Fujimoto}, {Henegar-Leon}, {Sneiderman}, {Tomida}, {Mori}, {Nakajima},
  {Terada}, {Holland}, {Loewenstein}, {Miller}, {Sawada}, {Kallman}, {Kaastra},
  {Done}, {Enoto}, {Bamba}, {Corrales}, {Ueda}, {Kara}, {Zhuravleva}, {Fujita},
  {Arai}, {Audard}, {Awaki}, {Ballhausen}, {Baluta}, {Bando}, {Behar},
  {Bialas}, {Boissay-Malaquin}, {Brenneman}, {Brown}, {Chiao}, {Cumbee}, {de
  Vries}, {den Herder}, {D{\'\i}az Trigo}, {DiPirro}, {Dotani}, {Carrero},
  {Ebisawa}, {Eckart}, {Eckert}, {Eguchi}, {Ezoe}, {Ferrigno}, {Foster},
  {Fukazawa}, {Fukushima}, {Furuzawa}, {Gallo}, {Garcia Martinez}, {Gorter},
  {Grim}, {Gu}, {Hagino}, {Hamaguchi}, {Hatsukade}, {Hayashi}, {Hayashi},
  {Hell}, {Hodges-Kluck}, {Horiuchi}, {Hornschemeier}, {Hoshino}, {Ichinohe},
  {Ikuta}, {Iizuka}, {Ishi}, {Ishida}, {Ishihama}, {Ishikawa}, {Ishimura},
  {Jaffe}, {Katsuda}, {Kanemaru}, {Kenyon}, {Kilbourne}, {Kimball}, {Kitamoto},
  {Kobayashi}, {Kohmura}, {Kubota}, {Leutenegger}, {Maeda}, {Markevitch},
  {Matsumoto}, {Matsuzaki}, {McCammon}, {McLaughlin}, {McNamara}, {Mernier},
  {Miko}, {Miller}, {Minesugi}, {Mitani}, {Mitsuishi}, {Mizumoto}, {Mizuno},
  {Mukai}, {Murakami}, {Mushotzky}, {Nakazawa}, {Natsukari}, {Ness}, {Nigo},
  {Nishiyama}, {Nobukawa}, {Nobukawa}, {Noda}, {Odaka}, {Ogawa}, {Ogawa},
  {Ogorzalek}, {Okajima}, {Okamoto}, {Ota}, {Ozaki}, {Paltani}, {Plucinsky},
  {Porter}, {Pottschmidt}, {Quero}, {Sasaki}, {Sato}, {Sato}, {Sato}, {Sato},
  {Seta}, {Shida}, {Shidatsu}, {Shigeto}, {Shipman}, {Shinozaki}, {Shirron},
  {Simionescu}, {Smith}, {Soong}, {Suzuki}, {Szymkowiak}, {Takahashi}, {Takeo},
  {Tamagawa}, {Tamura}, {Tanaka}, {Tanimoto}, {Terashima}, {Tsuboi},
  {Tsujimoto}, {Tsunemi}, {Tsuru}, {Uchida}, {Uchida}, {Uchida}, {Uchiyama},
  {Uno}, {Vink}, {Witthoeft}, {Wolfs}, {Yamada}, {Yamada}, {Yamaoka},
  {Yamasaki}, {Yamauchi}, {Yamauchi}, {Yanagase}, {Yaqoob}, {Yasuda},
  {Yoneyama}, {Yoshida}, \& {Yukita}}]{tashiro2025}
{Tashiro}, M., {Kelley}, R., {Watanabe}, S., {et~al.} 2025, \pasj, 77, S1,
  \dodoi{10.1093/pasj/psaf023}

\bibitem[{{Tchekhovskoy} {et~al.}(2011){Tchekhovskoy}, {Narayan}, \&
  {McKinney}}]{sasha2011}
{Tchekhovskoy}, A., {Narayan}, R., \& {McKinney}, J.~C. 2011, \mnras, 418, L79,
  \dodoi{10.1111/j.1745-3933.2011.01147.x}

\bibitem[{{The Xrism Collaboration} {et~al.}(2026){The Xrism Collaboration},
  {Audard}, {Awaki}, {Ballhausen}, {Bamba}, {Behar}, {Boissay-Malaquin},
  {Brenneman}, {Brown}, {Corrales}, {Costantini}, {Cumbee}, {D{\'\i}az Trigo},
  {Done}, {Dotani}, {Ebisawa}, {Eckart}, {Eckert}, {Eguchi}, {Enoto}, {Ezoe},
  {Foster}, {Fujimoto}, {Fujita}, {Fukazawa}, {Fukushima}, {Furuzawa}, {Gallo},
  {Garc{\'\i}a}, {Gu}, {Guainazzi}, {Hagino}, {Hamaguchi}, {Hatsukade},
  {Hayashi}, {Hayashi}, {Hell}, {Hodges-Kluck}, {Hornschemeier}, {Ichinohe},
  {Ishi}, {Ishida}, {Ishikawa}, {Ishisaki}, {Kaastra}, {Kallman}, {Kara},
  {Katsuda}, {Kanemaru}, {Kelley}, {Kilbourne}, {Kitamoto}, {Kobayashi},
  {Kohmura}, {Kubota}, {Leutenegger}, {Loewenstein}, {Maeda}, {Markevitch},
  {Matsumoto}, {Matsushita}, {McCammon}, {McNamara}, {Mernier}, {Miller},
  {Miller}, {Mitsuishi}, {Mizumoto}, {Mizuno}, {Mori}, {Mukai}, {Murakami},
  {Mushotzky}, {Nakajima}, {Nakazawa}, {Ness}, {Nobukawa}, {Nobukawa}, {Noda},
  {Odaka}, {Ogawa}, {Ogorzalek}, {Okajima}, {Ota}, {Paltani}, {Petre},
  {Plucinsky}, {Porter}, {Pottschmidt}, {Sato}, {Sato}, {Sawada}, {Seta},
  {Shidatsu}, {Simionescu}, {Smith}, {Suzuki}, {Szymkowiak}, {Takahashi},
  {Takeo}, {Tamagawa}, {Tamura}, {Tanaka}, {Tanimoto}, {Tashiro}, {Terada},
  {Terashima}, {Tsuboi}, {Tsujimoto}, {Tsunemi}, {Tsuru}, {T{\"u}mer},
  {Uchida}, {Uchida}, {Uchida}, {Uchiyama}, {Ueda}, {Uno}, {Vink}, {Watanabe},
  {Williams}, {Yamada}, {Yamada}, {Yamaguchi}, {Yamaoka}, {Yamasaki},
  {Yamauchi}, {Yamauchi}, {Yaqoob}, {Yoneyama}, {Yoshida}, {Yukita}, {Drury},
  {Hlavacek-Larrondo}, {Meunier}, {Migkas}, {Shefler}, {Stancil}, {Truong},
  {Ueda}, {Vigneron}, {Zuhone}, {Zhang}, {Heinrich}, {Zhuravleva}, \&
  {Bellomi}}]{xrism2026}
{The Xrism Collaboration}, {Audard}, M., {Awaki}, H., {et~al.} 2026, \nat, 650,
  309, \dodoi{10.1038/s41586-025-10017-x}

\bibitem[{{Vasudevan} \& {Fabian}(2007)}]{2007MNRAS.381.1235V}
{Vasudevan}, R.~V., \& {Fabian}, A.~C. 2007, \mnras, 381, 1235,
  \dodoi{10.1111/j.1365-2966.2007.12328.x}

\bibitem[{{Vaughan} {et~al.}(2003){Vaughan}, {Edelson}, {Warwick}, \&
  {Uttley}}]{Vaughanetal2003}
{Vaughan}, S., {Edelson}, R., {Warwick}, R.~S., \& {Uttley}, P. 2003, Monthly
  Notices of the Royal Astronomical Society, 345, 1271,
  \dodoi{10.1046/j.1365-2966.2003.07042.x}

\bibitem[{{Weaver} {et~al.}(2022){Weaver}, {Jorstad}, {Marscher}, {Morozova},
  {Troitsky}, {Agudo}, {G{\'o}mez}, {L{\"a}hteenm{\"a}ki}, {Tammi}, \&
  {Tornikoski}}]{Weaver2022}
{Weaver}, Z.~R., {Jorstad}, S.~G., {Marscher}, A.~P., {et~al.} 2022, Astrophys.
  J. Suppl. Ser., 260, 12, \dodoi{10.3847/1538-4365/ac58f1}

\bibitem[{{Wilms} {et~al.}(2000){Wilms}, {Allen}, \&
  {McCray}}]{2000ApJ...542..914W}
{Wilms}, J., {Allen}, A., \& {McCray}, R. 2000, \apj, 542, 914,
  \dodoi{10.1086/317016}

\bibitem[{{Yamazaki} {et~al.}(2013){Yamazaki}, {Fukazawa}, {Sasada}, {Itoh},
  {Nishino}, {Takahashi}, {Takaki}, {Kawabata}, {Yoshida}, \&
  {Uemura}}]{yamazaki2013}
{Yamazaki}, S., {Fukazawa}, Y., {Sasada}, M., {et~al.} 2013, \pasj, 65, 30,
  \dodoi{10.1093/pasj/65.2.30}

\end{thebibliography}
\bibliographystyle{aasjournal}

\renewcommand{\arraystretch}{1.5}

\begin{longtable}{lcc}
\caption{NGC 1275 flux measurements from Swift XRT observations. Fluxes are measured over the 0.3--10.0\,keV band. $1\sigma$ error ranges are reported. MJD represents the start time of the observation.} \\
\toprule
\textbf{OBS\_ID} & \textbf{MJD} & \textbf{Flux ($10^{-11}$ erg\,cm$^{-2}$\,s$^{-1}$)} \\
\midrule
\endfirsthead

\toprule
\textbf{OBS\_ID} & \textbf{MJD} & \textbf{Flux ($10^{-11}$ erg\,cm$^{-2}$\,s$^{-1}$)} \\
\midrule
\endhead

00036524001 & 54294.67348 & $3.288^{+0.055}_{-0.071}$ \\
00036524002 & 54440.68128 & $5.576^{+0.098}_{-0.114}$ \\
00030354003 & 55195.08536 & $3.747^{+0.083}_{-0.054}$ \\
00031770001 & 55399.87332 & $5.566^{+0.121}_{-0.159}$ \\
00031770002 & 55401.87800 & $5.861^{+0.142}_{-0.100}$ \\
00031770003 & 55403.55353 & $4.825^{+0.133}_{-0.144}$ \\
00031770004 & 55405.89494 & $4.990^{+0.121}_{-0.107}$ \\
00031770005 & 55407.69771 & $4.483^{+0.121}_{-0.093}$ \\
00031770006 & 55409.16855 & $4.218^{+0.111}_{-0.105}$ \\
00031770007 & 55411.10505 & $4.385^{+0.100}_{-0.076}$ \\
00031770008 & 55413.57692 & $4.870^{+0.135}_{-0.152}$ \\
00031770009 & 55415.05990 & $5.178^{+0.143}_{-0.106}$ \\
00031770010 & 55417.73560 & $4.661^{+0.101}_{-0.128}$ \\
00091128001 & 55747.13813 & $3.840^{+0.137}_{-0.174}$ \\
00091128002 & 55748.00250 & $4.002^{+0.179}_{-0.143}$ \\
00091128003 & 55749.21174 & $3.646^{+0.103}_{-0.112}$ \\
00091128004 & 55751.21868 & $3.717^{+0.106}_{-0.098}$ \\
00091128005 & 55752.09033 & $3.666^{+0.076}_{-0.077}$ \\
00032691001 & 56313.82473 & $4.431^{+0.349}_{-0.380}$ \\
00049799001 & 56386.46580 & $4.244^{+0.139}_{-0.218}$ \\
00049799002 & 56483.68244 & $3.458^{+0.282}_{-0.316}$ \\
00049799003 & 56485.95597 & $4.437^{+0.217}_{-0.194}$ \\
00049799004 & 56487.35796 & $4.233^{+0.046}_{-0.074}$ \\
00049799005 & 56499.96455 & $5.325^{+0.118}_{-0.115}$ \\
00049799006 & 56505.04979 & $4.746^{+0.117}_{-0.129}$ \\
00092034001 & 57064.35356 & $5.629^{+0.135}_{-0.139}$ \\
00092034002 & 57096.03512 & $4.871^{+0.132}_{-0.130}$ \\
00092034003 & 57228.09737 & $5.607^{+0.142}_{-0.149}$ \\
00092034004 & 57252.04383 & $4.663^{+0.135}_{-0.199}$ \\
00092034005 & 57281.55911 & $2.794^{+0.086}_{-0.131}$ \\
00081530001 & 57329.16671 & $4.808^{+0.069}_{-0.081}$ \\
00034380001 & 57437.86329 & $4.825^{+0.115}_{-0.057}$ \\
00034380002 & 57439.72359 & $4.536^{+0.093}_{-0.102}$ \\
00034380004 & 57441.39234 & $4.511^{+0.153}_{-0.100}$ \\
00034380005 & 57443.71888 & $3.623^{+0.101}_{-0.085}$ \\
00034380006 & 57444.71531 & $4.857^{+0.120}_{-0.114}$ \\
00034380007 & 57447.90033 & $4.722^{+0.134}_{-0.146}$ \\
00034380008 & 57449.43096 & $4.321^{+0.096}_{-0.143}$ \\
00034380010 & 57450.82320 & $4.466^{+0.109}_{-0.102}$ \\
00034380009 & 57451.02783 & $4.496^{+0.138}_{-0.097}$ \\
00034404001 & 57452.08721 & $4.792^{+0.117}_{-0.089}$ \\
00034380012 & 57453.81941 & $4.874^{+0.153}_{-0.123}$ \\
00034380013 & 57455.35516 & $4.570^{+0.082}_{-0.121}$ \\
00034380014 & 57457.21563 & $5.037^{+0.112}_{-0.104}$ \\
00034404002 & 57457.55827 & $4.520^{+0.205}_{-0.259}$ \\
00034380015 & 57459.00857 & $4.724^{+0.120}_{-0.129}$ \\
00034404003 & 57463.46257 & $4.235^{+0.142}_{-0.105}$ \\
00034765001 & 57691.02102 & $5.213^{+0.141}_{-0.148}$ \\
00034765002 & 57692.08656 & $5.542^{+0.163}_{-0.124}$ \\
00034765003 & 57693.12490 & $5.524^{+0.155}_{-0.144}$ \\
00034765004 & 57694.12195 & $5.024^{+0.139}_{-0.130}$ \\
00034765005 & 57695.11873 & $5.188^{+0.234}_{-0.204}$ \\
00034765006 & 57696.11430 & $5.869^{+0.132}_{-0.149}$ \\
00034765007 & 57697.11166 & $5.526^{+0.125}_{-0.124}$ \\
00034765008 & 57698.10791 & $5.410^{+0.163}_{-0.139}$ \\
00034765009 & 57699.10490 & $7.113^{+0.174}_{-0.128}$ \\
00034765010 & 57700.10149 & $6.016^{+0.158}_{-0.171}$ \\
00034765011 & 57701.09803 & $5.885^{+0.178}_{-0.182}$ \\
00034765012 & 57702.09370 & $5.883^{+0.135}_{-0.125}$ \\
00087312001 & 57753.00232 & $7.231^{+0.234}_{-0.192}$ \\
00087311001 & 57754.25442 & $7.962^{+0.283}_{-0.288}$ \\
00087311002 & 57755.18377 & $7.382^{+0.220}_{-0.261}$ \\
00087311003 & 57827.95858 & $5.419^{+0.213}_{-0.192}$ \\
00087312002 & 57833.19266 & $4.949^{+0.113}_{-0.151}$ \\
00087311004 & 57833.79750 & $5.086^{+0.219}_{-0.172}$ \\
00087311005 & 57836.51008 & $4.677^{+0.154}_{-0.142}$ \\
00087312003 & 57837.24887 & $4.445^{+0.255}_{-0.207}$ \\
00087312004 & 57838.04375 & $4.748^{+0.087}_{-0.077}$ \\
00087312005 & 57843.22897 & $4.961^{+0.121}_{-0.129}$ \\
03104795001 & 58454.44335 & $4.831^{+0.259}_{-0.287}$ \\
00034765013 & 59932.18550 & $8.269^{+0.152}_{-0.203}$ \\
00034765014 & 59933.91838 & $10.95^{+0.28}_{-0.22}$ \\
00034765016 & 59935.90493 & $8.666^{+0.226}_{-0.243}$ \\
00034765015 & 59936.24119 & $8.173^{+0.261}_{-0.184}$ \\
00034765017 & 59939.15502 & $6.428^{+0.245}_{-0.204}$ \\
00034765021 & 59955.89322 & $6.948^{+0.205}_{-0.182}$ \\
00034765023 & 59956.10606 & $11.34^{+0.33}_{-0.27}$ \\
00034765024 & 59956.89446 & $10.69^{+0.18}_{-0.19}$ \\
00034765026 & 59959.86633 & $7.520^{+0.194}_{-0.214}$ \\
00034765027 & 59975.58684 & $6.468^{+0.410}_{-0.376}$ \\
00034765031 & 59983.05737 & $6.881^{+1.285}_{-0.844}$ \\
00034765032 & 59985.71245 & $7.334^{+0.570}_{-0.608}$ \\
00034765039 & 60233.45275 & $4.708^{+0.209}_{-0.171}$ \\
00034765040 & 60234.45605 & $4.438^{+0.160}_{-0.193}$ \\
00034765041 & 60235.37001 & $6.027^{+0.417}_{-0.411}$ \\
00034765042 & 60290.15940 & $6.419^{+1.418}_{-1.420}$ \\
00034765043 & 60291.21633 & $5.067^{+0.231}_{-0.227}$ \\
00019048001 & 60699.92675 & $6.235^{+0.205}_{-0.129}$ \\
03000223001 & 61008.50517 & $4.989^{+0.187}_{-0.160}$ \\
\bottomrule
\end{longtable}

\begin{figure}[t]
   \centering
   \includegraphics[width=1.0\columnwidth]{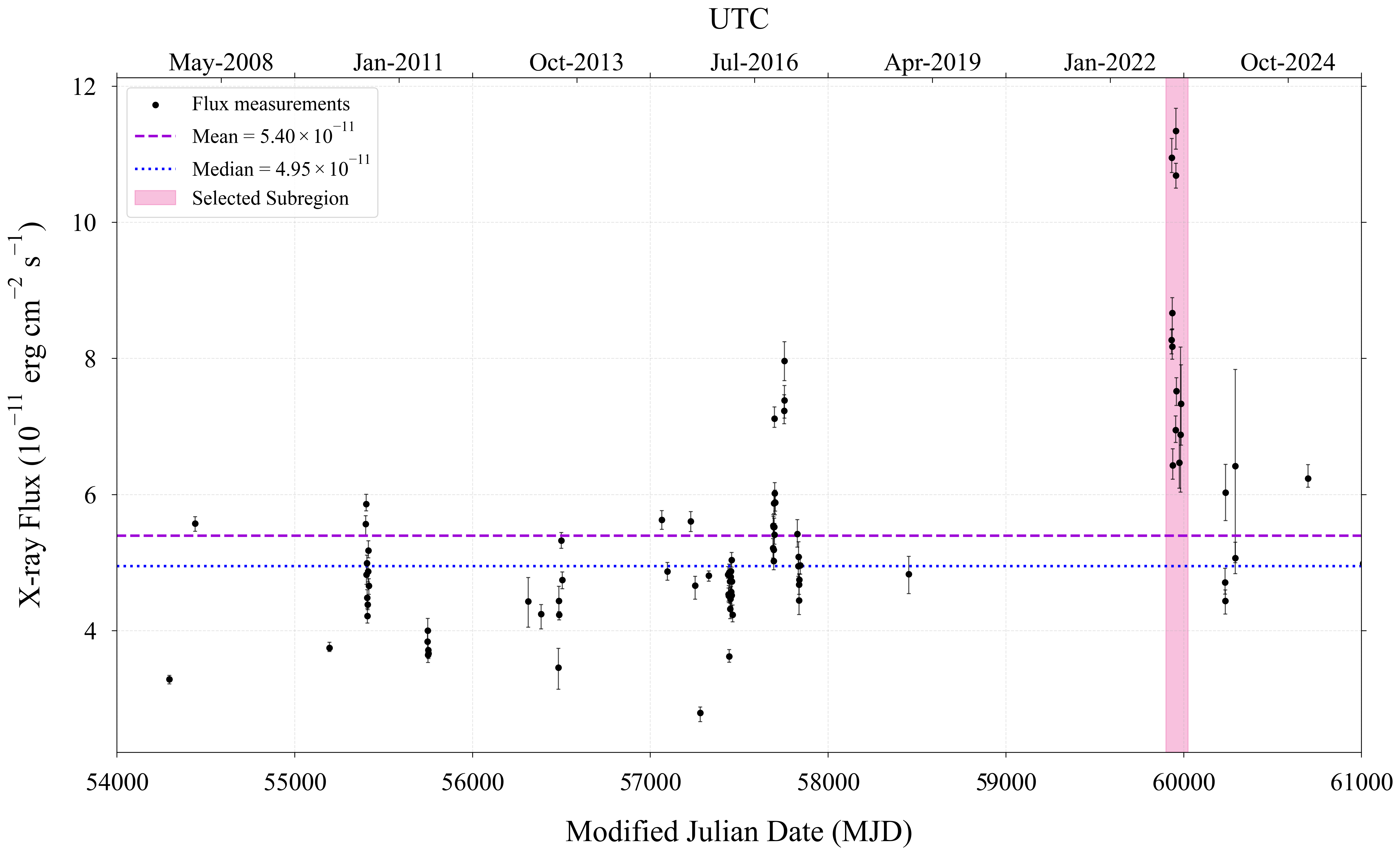}
   \caption{NGC 1275 X-ray flux (in $\text{ erg cm}^{-2} \text{ s}^{-1}$) versus time (in MJD). $1\sigma$ flux errors are shown. The purple line corresponds to the mean flux value over this time period, and the blue line corresponds to the median flux during the same time period. The large flare around MJD 60000 is highlighted in pink.}
  \label{fig:fek}
\end{figure}

\begin{figure}[t]
   \centering
   \includegraphics[width=0.49\columnwidth]{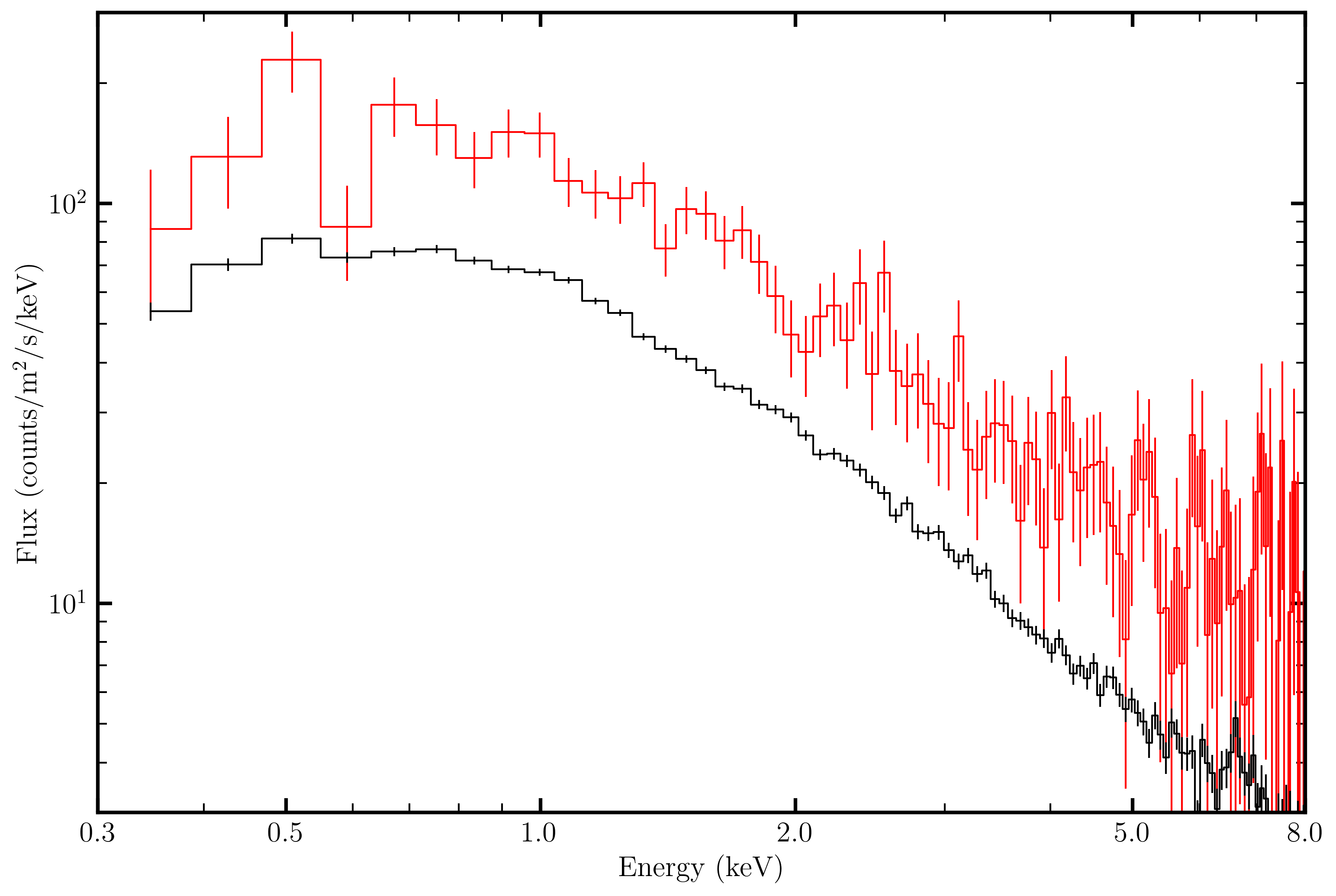}
   \includegraphics[width=0.49\columnwidth]{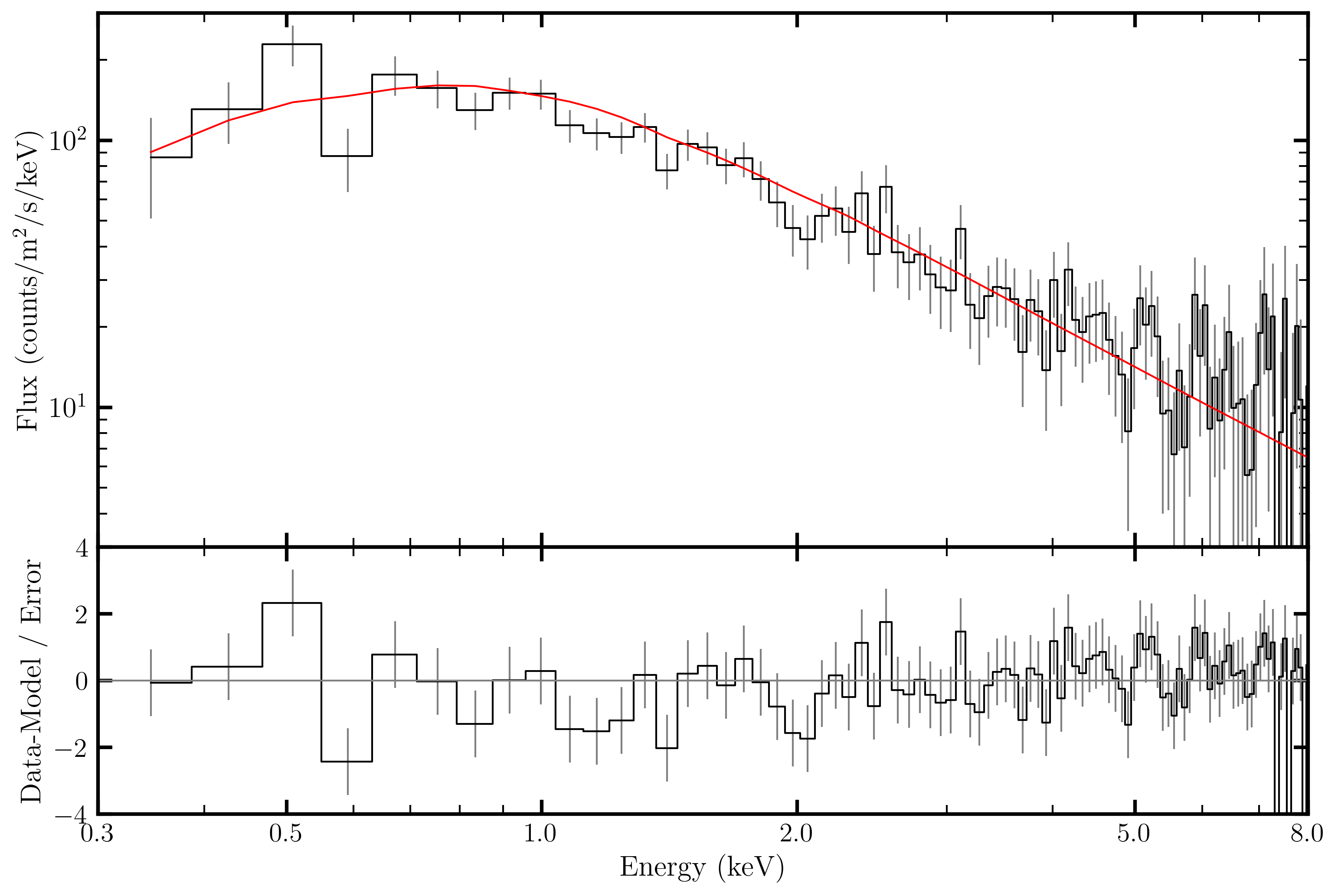}
   \caption{Left: Swift/XRT spectra of the Perseus Cluster and NGC~1275, illustrating variability in the AGN.  A time-averaged spectrum is shown in black.  The brightest single observation is shown in red (ObsID 34765023).  Right: The difference spectrum (ObsID 34765023 minus the time-averaged spectrum), fit with an absorbed $\Gamma = 1.7$ power-law over the band with sufficient signal.}
  \label{fig:fek}
\end{figure}

\begin{figure}[t]
   \centering
   \includegraphics[width=1.0\columnwidth]{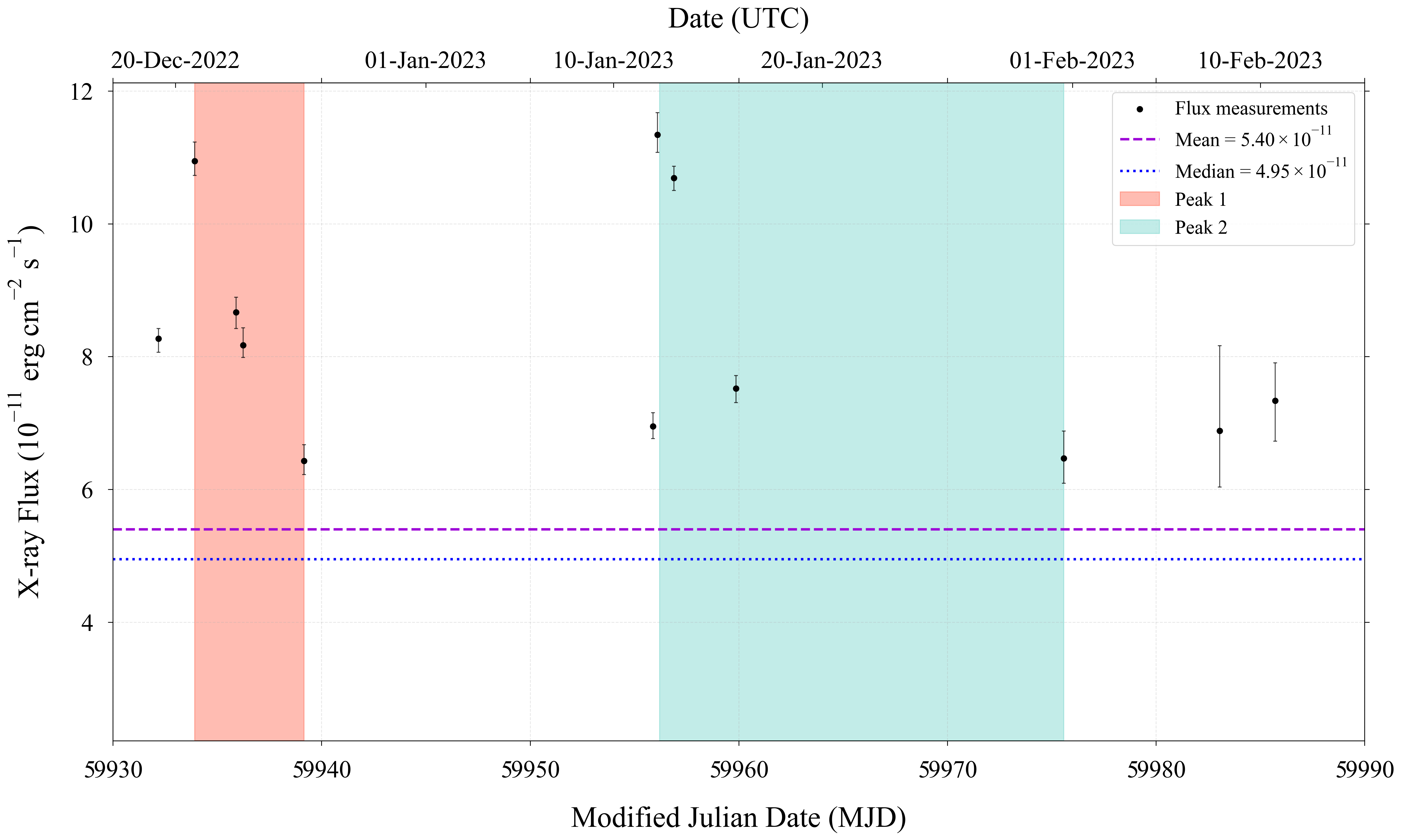}
   \caption{Zoomed-in view of the peak at 60000 MJD highlighted in Fig. 1. The large peak at 60000 MJD is composed of two smaller subregions, which are highlighted in red and blue, respectively. Each of the two peaks was analyzed for similarities to the known behavior of a tidal disruption event. The highlighted regions span the maximum to minimum values of the decay regions of each peak.  The graph still displays NGC 1275 X-ray flux (in $\text{ erg cm}^{-2} \text{ s}^{-1}$) versus time (in MJD). The purple line corresponds to the mean flux value, and the blue line corresponds to the median flux.}
  \label{fig:fek}
\end{figure}

\begin{figure}[t]
   \centering
   \includegraphics[width=1.0\columnwidth]{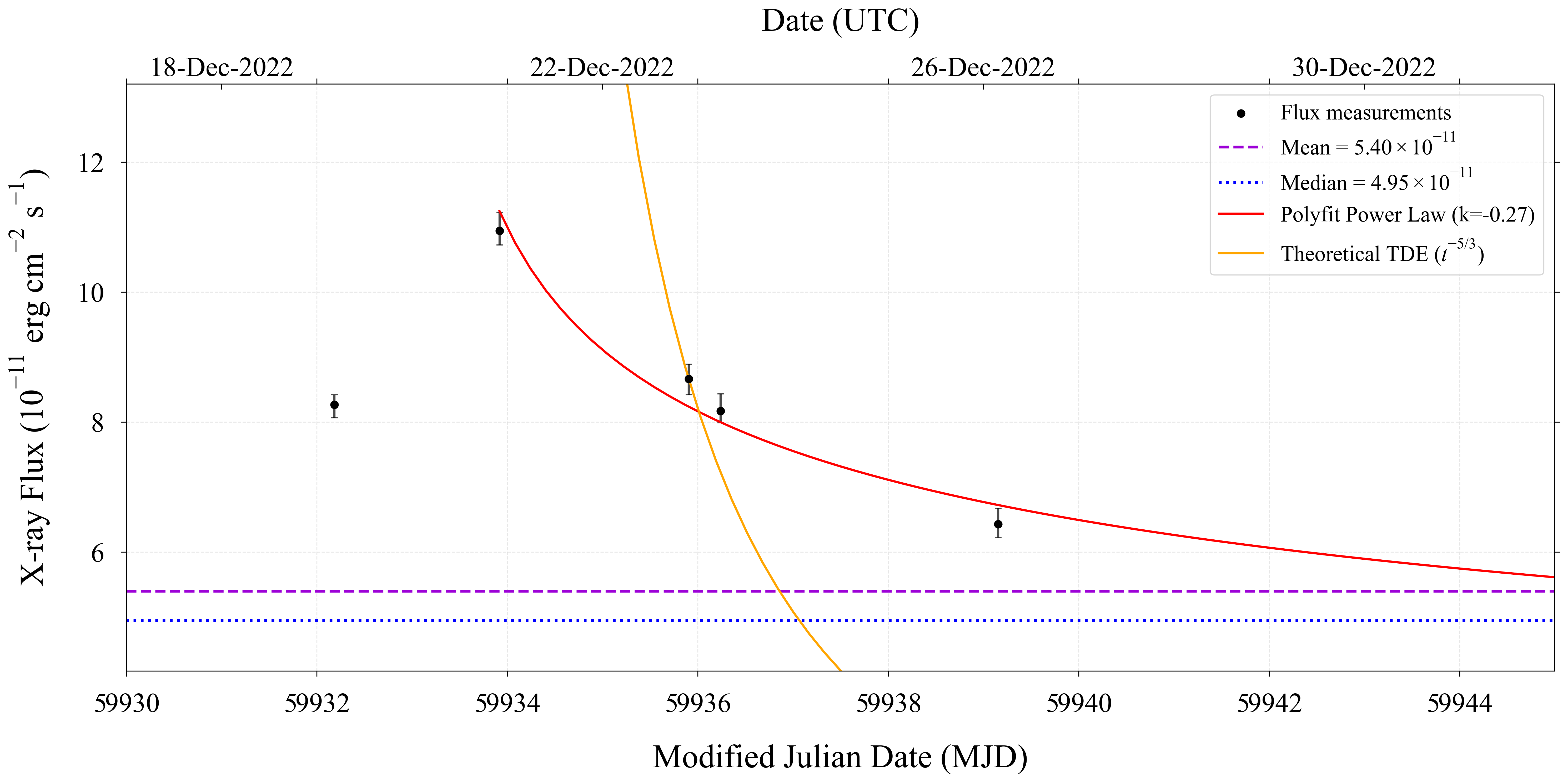}
   \caption{Close-in look at the first subregion (the red region of Fig. 3) of the large flare (highlighted in Fig. 1). We attempted to fit the data as a tidal disruption event (TDE) by using a polyfit power law and comparing to a theoretical model. The red line corresponds to the calculated polyfit, and the orange line corresponds to the theoretical shape of the data if the data was a TDE. Notably, the fit to the data points within this peak does not resemble a TDE.}
  \label{fig:fek}
\end{figure}

\begin{figure}[t]
   \centering
   \includegraphics[width=1.0\columnwidth]{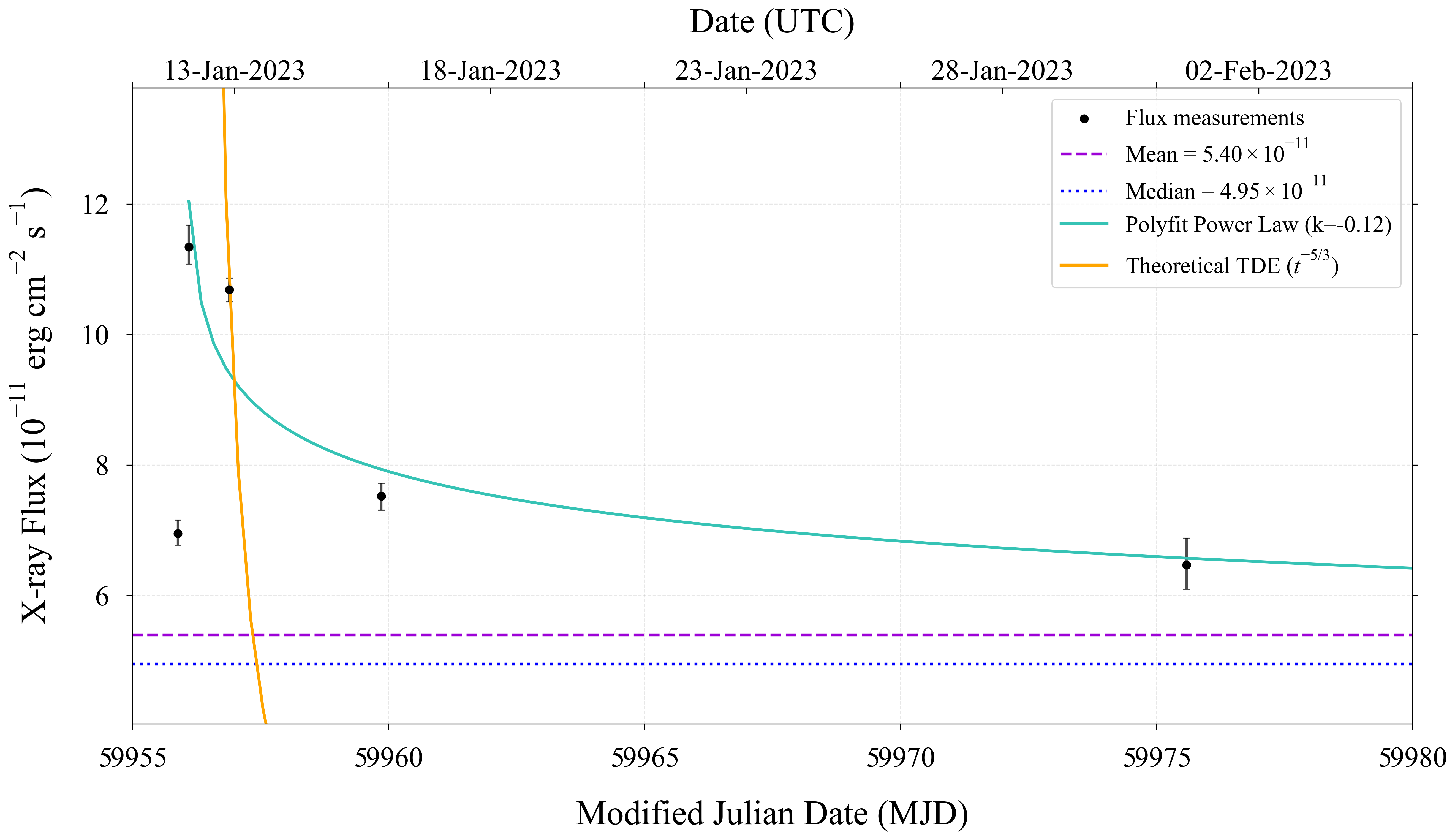}
   \caption{Close-in look at the second subregion (the blue region of Fig. 3) of the large flare highlighted in pink in Fig. 1. We again attempted to fit the data as a TDE by using a polyfit power law and comparing to a theoretical model. The blue line corresponds to the calculated fit, and the orange line corresponds to the theoretical fit of the data if the data was a TDE. Once again, the fit to this peak does not resemble the theoretical decay pattern of a TDE.}
  \label{fig:fek}
\end{figure}

\begin{figure}[t]
   \centering
   \includegraphics[width=1.0\columnwidth]{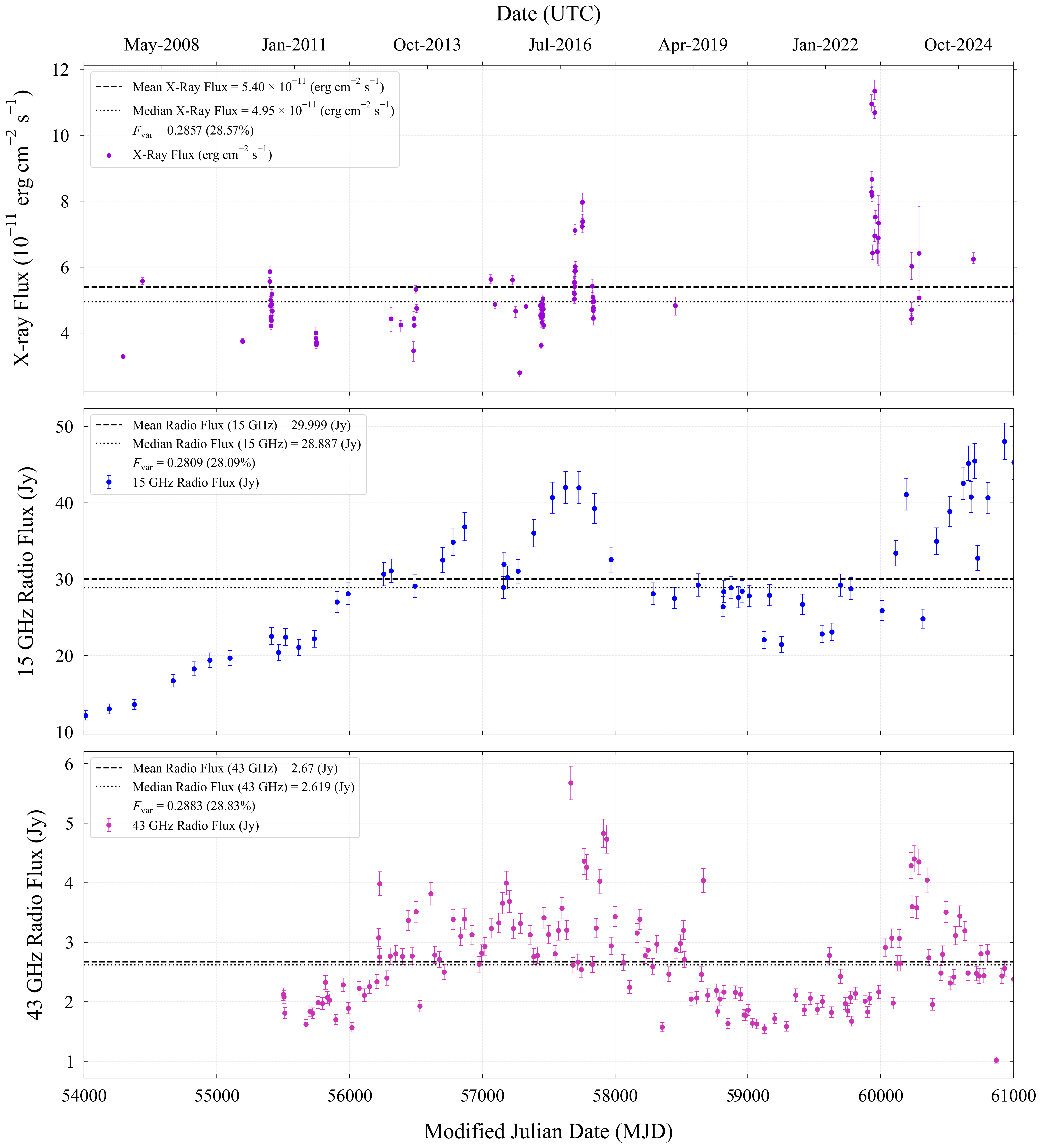}
   \caption{The Swift/XRT, MOJAVE, and VLBA-BU-BLAZAR light curve of NGC~1275 over the same time span. The MOJAVE data are based on observations made using the VLBA at 15~Ghz \citep{lister2018}. The VLBA-BU-BLAZAR data is based on observations made using the VLBA at 43~GHz \citep{Weaver2022}. The X-ray and radio data have roughly the same fractional variability over similar time frames. The X-ray flare near to MJD 60000 does not have a clear counterpart in the radio data, but it is notable that contemporaneous radio flux points are below the local mean and median values.  The subsequent increase in radio flux density over $\sim1000$~days could be tied to the X-ray flare, if the flare marked a sudden increase in the mass accretion rate onto the black hole that subsequently propagated into the jet.}
  \label{fig:radio}
\end{figure}

\end{document}